**Adam SIEMIEŃSKI**[1], **Alicja KEBEL**[1]**, Piotr KLAJNER**[2]
[1]Department of Biomechanics, University School of Physical Education in Wrocław
[2]Institute of Physics, Wrocław University of Technology

# FATIGUE INDEPENDENT AMPLITUDE-FREQUENCY CORRELATIONS IN EMG SIGNALS

Summary. In order to assess fatigue independent amplitude-frequency correlations in EMG signals we asked nineteen male subjects to perform a series of isometric muscular contractions by extensors of the knee joint. Different amplitudes of the signal were due to randomly varying both the joint moment and the overall amplification factor of the EMG apparatus. Mean and median frequency, RMS and mean absolute value were calculated for every combination of joint moment and amplification at the original sampling rate of 5 kHz and at several simulated lower sampling rates. Negative Spearman and Kendall amplitude-frequency correlation coefficients were found, and they were more pronounced at high sampling rates.

## 1. INTRODUCTION

Surface electromyography (SEMG) signals are usually processed by using some data reduction techniques to obtain quantities describing their amplitude and dominating frequency [1]. The four most often used are RMS and MAV (mean absolute value) for amplitude, and median frequency ($F_{med}$) and mean frequency ($F_{mean}$) for frequency. All four are influenced by both the level of produced muscular force and fatigue. It is also commonly accepted in the literature that fatigue can be deduced from SEMG signal behavior only when simultaneously a drop of $F_{med}$ or $F_{mean}$ and an increase of signal amplitude is observed [2, 3]. In fact, when force produced by a muscle (or moment of force produced at a joint) is kept unchanged with time, fatigue results in a downward shift of the frequency spectrum of the signal and an increase of signal amplitude [2, 3]. That is why the extent or onset of fatigue is typically assessed by simultaneous monitoring of frequency content and amplitude of surface EMG signals collected from engaged muscles and not just the frequency parameters alone [1].

However, such negative amplitude-frequency correlations may occur also at no fatigue due to the varying signal to noise ratio, which could blur the interpretation of EMG signal frequency shifts in terms of fatigue. The objective of this work is therefore to perform a series of EMG measurements under conditions leading to a wide range of signal amplitudes but minimizing the effects of fatigue, and to assess the resulting relationships between amplitude and frequency measures of the EMG signal.

## 2. MATERIAL AND METHOD

Nineteen healthy male volunteers, students of the University School of Physical Education in Wrocław, took part in the experiment. Mean values and standard deviations of their age, body height and body mass are gathered in Table 1.

A. Siemieński, A. Kebel, P. Klajner

EMG signals were collected by a pair of surface electrodes placed over the vastus medialis head of the quadriceps muscle, between its motor point and tendon, collinearly with muscle fibers. The signals were processed by the Octopus SEMG device: two stage amplification including a preamplifier (ten times) placed just a few centimeters from the electrode site and the main amplifier (with adjustable amplification) resulting in an overall amplification factor that could be varied between 100 and 2000 times. The amplified signals were sampled at a rate of 5000 Hz and recorded in digital form on the hard disk of a PC computer.

**Table. 1 Characteristics of subjects – mean values and standard deviations**

| age [years] (± SD) | body height [m] (± SD) | body mass [kg] (± SD) |
|---|---|---|
| 20,29 (± 0,47) | 1,80 (± 0,04) | 74,94 (± 8,01) |

The subjects were asked to assume an upright sitting position on a measuring stand which included a stiff chair with adjustable back support. The position of their right lower limb was adjusted so as to place the knee joint transverse axis as collinearly as possible with the axis of the torque measuring device, and this was done for the knee angle of 90º. The subjects were then asked to try to extend their right limb at the knee joint by pushing against a fixed lever of the torque measuring device placed just above the ankle joint, which resulted in isometric activity of the extensors of the knee joint. After a short warm-up, each subject was first asked to produce his maximum isometric knee extension moment, corresponding to maximum voluntary contraction (MVC) of the quadriceps muscle. The maximum moment of force, produced at MVC, was recorded and later used as reference for all the remaining trials. Then, the main session started which consisted of two-second-long trials realized with visual feedback at 5%, 10%, 20%, 30% and 50%, of MVC, and at the overall amplification factor of the apparatus equal to 100, 200, 500, 1000, 2000. The sequence of torque levels and amplification factors was random to minimize the effects of fatigue. Also, if a subject reported fatigue at some point of the session a rest break of a few minutes was provided. At the end of the session the knee joint moment at quadriceps MVC was measured again as another test of possible presence of fatigue.

For each subject, the experimental session was repeated on the next day. To enhance repeatability of the results, the EMG electrodes were then placed at the sites marked during the first session.

Two measures of the amplitude of the EMG signal, RMS and mean absolute value (MAV), and two measures of the dominating signal frequency, the mean frequency ($F_{mean}$) and the median frequency ($F_{med}$), were calculated for each two-second-long trial. These four quantities were evaluated both for the original signals sampled at the rate of 5 kHz and for a number of lower sampling rates simulated by decimating the original data points by integer factors. Spearman and Kendall correlation coefficients were calculated between the measures of amplitude and frequency for some of those sampling rates.

3. RESULTS

The measurement results for one subject and one session consist of four numbers representing RMS, MAV, $F_{mean}$, $F_{med}$ calculated for each (out of five) muscle force level, each (out of five) overall amplification factor and each (out of one hundred: 5000Hz, 2500Hz, 1250Hz, …, 50Hz) sampling rate. Out of these four quantities the two amplitude measures behaved quite differently from the frequency measures as the sampling rate increased. The EMG signal amplitude did not change with sampling rate, except for sampling rates below 250 Hz, where a chaotic behavior could be seen (Fig. 1. brings a representative example of such a behavior for RMS; the other amplitude measure – MAV – behaved similarly).

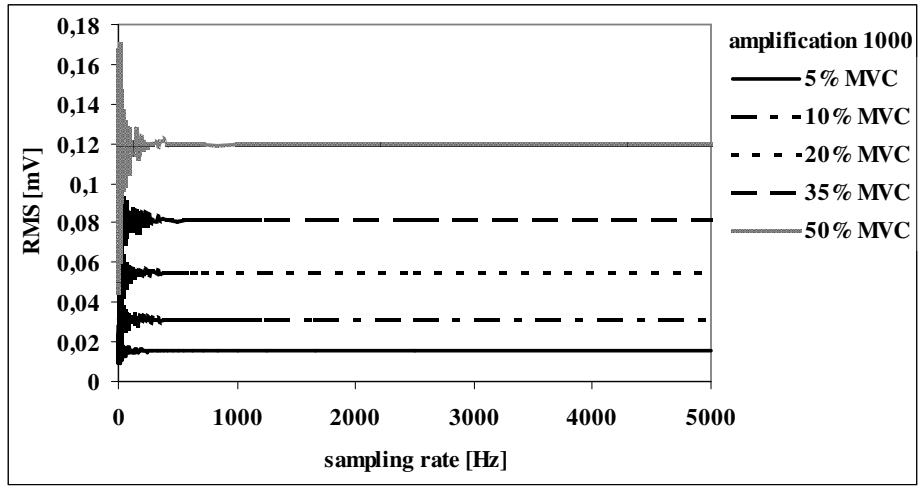

**Fig. 1 Dependence of the RMS value of vastus medialis EMG signal on muscle force level and on sampling rate**

Contrasting with the above amplitude – sampling rate relationship was its counterpart for the two estimates of the dominating signal frequency. There was no chaotic behavior, even at very low sampling rates, in the dependence of the signal median frequency on sampling rate (Fig. 2), and all the five curves corresponding to knee joint extending moments equal to 5%-50% of MVC showed a consistent growth, which was more pronounced at lower muscle forces.
As for the mean frequency of the EMG signal, its dependence on sampling rate was qualitatively similar to that shown by the median frequency but the growth rates were generally higher and not vanishing even at the highest muscle forces investigated, i.e. 50 % MVC (Fig. 3).

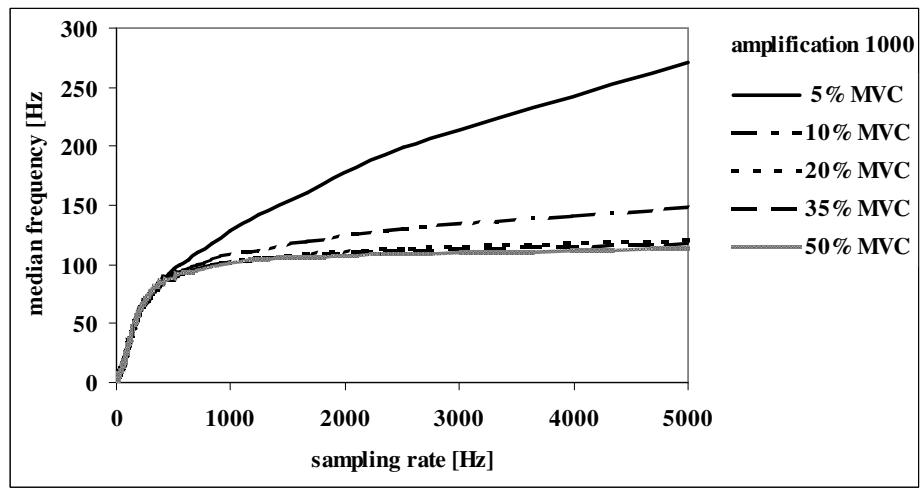

**Fig. 2 Dependence of the median frequency of vastus medialis EMG signal on muscle force level and on sampling rate**

A. Siemieński, A. Kebel, P. Klajner

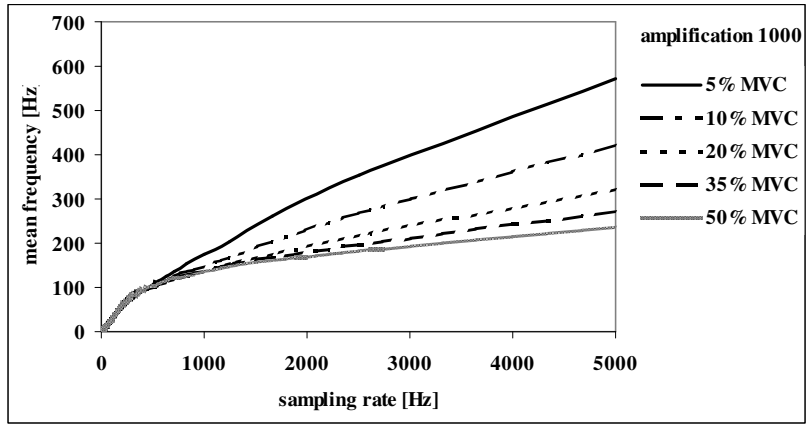

**Fig. 3 Dependence of the mean frequency of vastus medialis EMG signal on muscle force level and on sampling rate**

Interestingly enough, even though the two frequency measures generally increased as sampling rate increased, their behavior was independent of the muscle force produced until a sampling rate of about 500 Hz was reached. Indeed, in Fig. 2 and Fig. 3 the bunches of curves corresponding to different force levels start as essentially one curve and then split at this frequency, giving rise to a possible relationship between amplitude and frequency measures, the more pronounced the more split the curves are. Above 500 Hz, where the five curves become more and more split, an increase in signal amplitude (i.e. moving vertically across a bunch of curves from the 5% curve to the 50% curve) should result in a decrease of the median or mean frequency. That is why negative correlation between amplitude measures and frequency measures of EMG signal should be expected for sampling rates higher than 500 Hz. Another illustration of this effect is proposed in Fig. 4. where the dependence of the mean

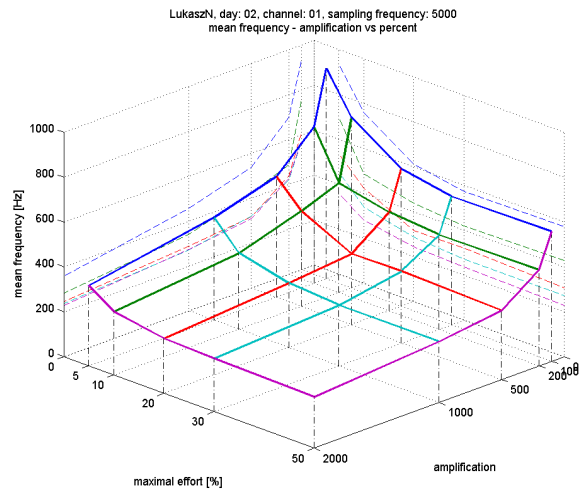

**Fig. 4 Dependence of the mean frequency of vastus medialis EMG signal on muscle force level and on the overall amplification factor at the highest sampling rate (5kHz)**

frequency on force level and amplification factor is depicted as a surface graph. The height of the surface graph shows the value of the mean frequency for a given force level and a given amplification factor. At this high sampling rate, 5 kHz, the mean frequency of the EMG signal increases rapidly as the force level or the amplification factor (or both) decreases, which, again, can be thought of as a negative amplitude-frequency correlation.

In order to assess this correlation the results were grouped into subsets containing data corresponding to a given amplification factor. Within each of those subsets were the results of calculation of the two amplitude measures and the two frequency measures found in all the subjects at five different force levels and at one hundred different sampling rates. All the subsets contained the same amount of data and they differed only by the overall amplification factor of the EMG apparatus. Correlation analysis performed within those subsets of data showed that the amplitude and frequency measures of the EMG signals were not correlated for low sampling rates but as the sampling rate increased passed the critical value of 500 Hz they became more and more interdependent.

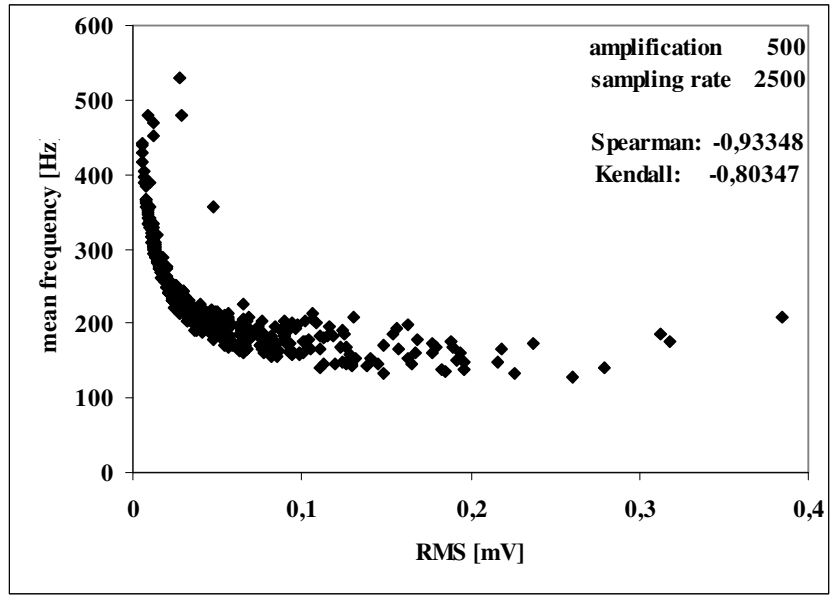

**Fig. 5 Mean frequency – RMS relationship for the overall amplification factor 500 and sampling rate 2500Hz – all subjects**

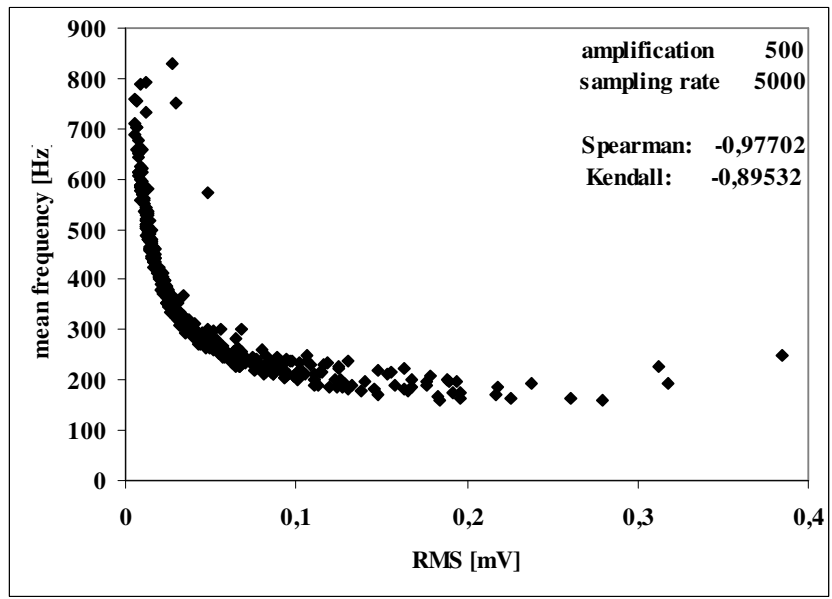

**Fig. 6 Mean frequency – RMS relationship for the overall amplification factor 500 and sampling rate 5000Hz – all subjects**

Two examples illustrating this tendency of amplitude-frequency correlation to become more and more evident as sampling rate increased are shown in Fig. 5. and Fig. 6. In the first of these two figures the interdependence is shown between the mean frequency of the EMG signal and its RMS value, at the overall amplification factor 500 and sampling rate 2500 Hz. Most data points lie on or near a hyperbola and Spearman and Kendall correlation coefficients between mean frequency and RMS are -0.93348 and -0.80347, respectively. For data corresponding to sampling rate 5 kHz, they become -0.97702 and -0.89532, respectively, and the data points follow a hyperbola even more closely (Fig. 6).

4. CONCLUSIONS

Our experimental protocol was designed to produce a wide variety of stationary EMG signals differing as much as possible by their amplitude, and to keep fatigue influence to a minimum. We looked at the behavior of measures of amplitude and measures of dominating frequency of the signals at sampling rates ranging from 50 Hz to 5 kHz. Our main conclusions are:
- amplitude measures such as RMS and MAV do not depend on sampling rate
- dominating frequency measures increase as sampling rate increases
- dominating frequency measures do not depend on signal amplitude for sampling rates lower than 500 Hz; for higher sampling rates, the smaller signal amplitude the more steeply the frequency measure increases with sampling rate
- this gives rise to negative amplitude-frequency correlations that occur at sampling rates higher than 500 Hz and become more pronounced as sampling rate increases
- such correlations, found here in an experiment designed to minimize fatigue, could shadow the response of amplitude and frequency measures to fatigue and should therefore be taken into account when monitoring the onset and development of fatigue
- to minimize these effects oversampling of EMG signals should be avoided